\documentclass[a4paper, 11pt]{article}
\usepackage{jheppub} 
\usepackage{lineno}
\usepackage{hyperref}
\usepackage{stackengine}
\usepackage{braket}
\usepackage{graphicx}
\usepackage{lipsum} 
\usepackage{float} 

\title{\boldmath Benchmarking quantum chaos from geometric complexity}

\author[1]{Arpan Bhattacharyya,}
\affiliation[1]{Department of Physics, Indian Institute of Technology, Gandhinagar,
Gujarat-382355, India}
\emailAdd{abhattacharyya@iitgn.ac.in}

\author[2]{Suddhasattwa Brahma,}
\affiliation[2]{Higgs Centre for Theoretical Physics, School of Physics \& Astronomy,\\ University of Edinburgh, Edinburgh EH9 3FD, Scotland, UK}
\emailAdd{suddhasattwa.brahma@gmail.com}

\author[3,4]{Satyaki Chowdhury,}
\affiliation[3]{Institute of Theoretical Physics, Jagiellonian University, Lojasiewicza 11, 30-348 Cracow, Poland}
\emailAdd{satyaki.chowdhury@doctoral.uj.edu.pl}
\affiliation[4]{Doctoral School of Exact and Natural Sciences, Jagiellonian University, Lojasiewicza 11, 30-348 Cracow, Poland}

\author[2]{and Xiancong Luo}
\emailAdd{x.luo-35@sms.ed.ac.uk}

\abstract{Recent studies have shown that there is a strong interplay between quantum complexity and quantum chaos. In this work, we consider a new method to study geometric complexity for interacting non-Gaussian quantum mechanical systems to benchmark the quantum chaos in a well-known oscillator model. In particular, we study the circuit complexity for the unitary time-evolution operator of a non-Gaussian bosonic quantum mechanical system.  Our results indicate that, within some limitations, geometric complexity can indeed be a good indicator of quantum chaos.}

\begin{document}
\maketitle
\flushbottom

\section{Introduction}

Chaos is a ubiquitous phenomenon in the macroscopic world. Unlike random systems, chaotic systems come from deterministic theories. However, due to the lack of arbitrarily precise measurements, the exponential sensitivity to initial conditions in chaotic systems makes long-term prediction unrealistic. Typically, one denotes the timescale for validity of deterministic evolution as the `Lyapunov time'. Classically\footnote{By classical, we would typically mean Newtonian dynamics or General Relativistic systems.}, there are several well-behaved criterion to distinguish chaotic systems without fully having to analyse their dynamics. Firstly, for chaotic systems, two nearby trajectories in phase space start separating exponentially according to Lyapunov exponents while for non-chaotic ones, the distance between trajectories evolves according to some power law. Secondly, for a bounded system, the phase space of chaotic system often exhibit the existence of a strange attractor behaviour\footnote{An attractor is a set of points in phase space that the system evolves towards over time regardless of its initial starting point.}, which has a fractal (non-integer) dimension. Mathematically, we can also determine the integrability of a Hamiltonian system from its phase-space dynamics, which is a sufficient (although not necessary) condition for being non-chaotic \cite{DAlessio:2015qtq}. 

On the other hand, we now (quite rightly) assume that the world around us is fundamentally quantum in nature. However, the quantum analogue of the classical chaos described above is yet not that obvious. The minimal unit in phase space for quantum mechanical systems is not a point anymore because of the Heisenberg uncertainty principle, and therefore the analogy with the trajectory separation is not available any longer. In addition, quantum states that differ slightly in their initial conditions will maintain their inner product throughout the entire evolution due to unitarity, and hence this is not a good measure to quantify the ``growth'' in the separation of these states. Moreover, since we understand the world through classical instruments, it implies that we do need a correspondence between quantum microphysics and macroscopic observables. This, of course, includes the classical/quantum correspondence for chaos. Generally, quantum chaos refers to the study of the sub-class of quantum systems whose classical limits are chaotic. Going beyond the pivotal role it plays in quantum many-body physics, recently quantum chaos has also had increasingly closer interactions with quantum gravity \cite{Shenker:2013pqa,Shenker:2013yza,Shenker:2014cwa,Roberts:2014isa,Maldacena:2015waa,Dittrich:2016hvj,Turiaci:2016cvo,Polchinski:2016xgd,Jensen:2016pah,Stanford:2019vob,Hashimoto:2021afd, Weber:2024ieq, Brahma:2024sie,DeFalco:2020yys,DeFalco:2021uak}. For example, in Holography, it is argued that black holes are described by maximally chaotic (dual) systems such as the Sachdev-Ye-Kitaev (SYK) model, which imposes a bound on the quantum Lyapunov exponent: $\lambda \leq 2 \pi k_{\mathrm{B}} T / \hbar$ \cite{Maldacena:2015waa}.

There are a few proposals which have been advanced to diagnose quantum chaos, namely Random Matrix Theory (RMT) \cite{Guhr:1997ve,DAlessio:2015qtq}, Out-of-Time-Ordered Correlations (OTOCs) \cite{larkin1969quasiclassical}, Entangelment Entropy \cite{PhysRevE.86.010102, PhysRevE.87.042135}, Eigenstate Thermalization Hypothesis (ETH) \cite{Srednicki:1994mfb}, Berry's Phase \cite{berry2001chaos} and fractal methods \cite{Cornish:1996hx,Bojowald:2023fas}. These techniques have been widely used and they vary in their ability to successfully characterize features of a quantum chaotic system. However, not only are some of these methods very difficult to compute, they also have limitations in identifying chaos in complicated setups. For example, recent studies on the mass-deformed SYK model revealed a conflict between the RMT and OTOC diagnostics \cite{Nosaka:2018iat,Hashimoto:2017oit}, and it was argued that the two probes capture features for two different temporal regimes. While OTOC could only capture the early-time behaviour of the system \cite{bhattacharyya2021multi}, RMT was successful in determining the late-time dynamics. 

Motivated by the need to have a better understanding of quantum chaos and the lack of a universal, computable probe to diagnose it, \textit{quantum complexity} was recently proposed as a complementary diagnostic which might be able to do so \cite{Bhattacharyya:2020art,Bhattacharyya:2021cwf,Bhattacharyya:2019txx,Balasubramanian:2019wgd,Parker:2018yvk,Nandy:2024htc,Brown:2016wib,Brown:2017jil,Auzzi:2020idm}. In this paper, we explore the effectiveness of the notion of complexity to identify the dynamics of a quantum chaotic systems. If we assume that the universe is a quantum computer with a certain capacity, circuit complexity measures how difficult it is to prepare a state starting from another quantum state \cite{https://doi.org/10.48550/arxiv.quant-ph/0502070, Nielsen_2006,https://doi.org/10.48550/arxiv.quant-ph/0701004, Jefferson:2017sdb,Bhattacharyya:2019kvj}. On the other hand, in a quantum chaotic system, the effect of a quench acting on a state is expected to be quickly scrambled across the system. Due to large number of basis states, it is not realistic to describe the fine-grained time-evolution of the system, \textit{i.e.,} the one given by the unitary evolution in terms of Hamiltonian. Instead, a coarse-grained measure needs to be considered. The time dependence of complexity in some sense indicates how fast (or how difficult) is this scrambling process by defining another measure of evolution. In other words, chaos may be related to the potential of minimal evolution instead of physical evolution. In this sense, complexity may serve as a diagnosis to detect quantum chaos. By now, there is a quite a few papers that have described how circuit complexity can be used to diagnose chaos \cite{Ali:2019zcj, Bhattacharyya:2020art,Yang:2019iav,Balasubramanian:2021mxo,Bhattacharyya:2020iic,Bhattacharyya:2019txx, Bhargava:2020fhl}. In particular, the early evolution of complexity for a chaotic system, with $N$ degrees of freedom, has been claimed to show linear growth \cite{Bhattacharyya:2023grv} (until the time $t\sim e^N$ when the complexity saturates).  

There are different notions of complexity. Krylov complexity is one such that quantifies the operator growth through its dependence on the so-called `Lanczoz coefficients'\footnote{For computing Krylov complexity, the evolution of operator growth is mapped to a single particle hopping on an one-dimensional chain with the hopping amplitudes governed by the Lanczos coefficients \cite{Parker:2018yvk, Caputa:2021sib}.} \cite{Parker:2018yvk, viswanath1994recursion, Dymarsky:2019elm}. Since one expects chaotic systems to show complex growth of local operators in time, it has been shown how the statistics of these Lanczos coefficients can be related to quantum chaos \cite{Balasubramanian:2022dnj, Hashimoto:2023swv, Bhattacharyya:2023dhp}. Krylov complexity has been studied recently for well-known chaotic systems, both with \cite{Hashimoto:2023swv} and without \cite{Rabinovici:2022beu,Chapman:2024pdw} having classical counterparts\footnote{For more details, interested readers are referred to the following review \cite{Nandy:2024htc}, and references therein.}.  In this work, we shall use a different measure of complexity, namely geometric (circuit) complexity \cite{Nielsen_2006,https://doi.org/10.48550/arxiv.quant-ph/0502070,https://doi.org/10.48550/arxiv.quant-ph/0701004,Jefferson:2017sdb,Chapman:2017rqy,Khan:2018rzm,Hackl:2018ptj,Guo:2018kzl, Bhattacharyya:2018bbv, Ali:2018fcz,Chapman:2018hou}, as the tool to diagnose chaos in a coupled oscillator system. We will make the definition of geometric complexity more precise later on, but for now, let us point out why our work is a significant improvement upon previous attempts made in this direction. 

In \cite{Ali:2019zcj}, a concrete way to diagnose quantum chaos from geometric complexity had been proposed. In order to diagnose chaos in quantum systems, one proposal is to use the complexity between the reference state and target state by evolving the reference state forward in time $t$ with Hamiltonian $H$ and backwards in time $t$ with Hamiltonian $H+\delta H$, where $\delta H $ is the small perturbation, that is:
\begin{equation}
\left|\psi_{\mathrm{T}}\right\rangle=e^{i (\hat{H}+ \hat{\delta H} )t} e^{-i \hat{H} t}\left|\psi_{\mathrm{R}}\right\rangle:=U\left|\psi_{\mathrm{R}}\right\rangle \,.\label{perturb}
\end{equation}
The argument is that a linear growth of complexity will mean that the small perturbation grows in time and can be considered to characterise quantum chaos in the system. Although this idea is certainly a novel one, it has one major loophole. A simple check of oscillator systems with a quadratic potentials shows that the linear growth does not always refer to quantum chaos. If the classical potential has an unstable saddle point, the complexity transfers from showing an oscillating behaviour to having linear growth. This is best reflected in the case of an inverse harmonic oscillator, which is a classically solvable system with no chaotic behaviour, and yet has linear growth of complexity. The growth comes from the fact that the distance of trajectories that begin very close from nearby saddle points will in general grow exponentially, leading to a false Lyapunov parameter.  In fact, this loophole in using complexity as a way to  detect chaos is the same one that appears when using Lyapunov coefficients of OTOCs, which grow exponentially in integrable systems containing isolated saddle points \cite{Xu:2019lhc}. On the other hand, geometric complexity is difficult to compute for non-Gaussian setups which do not have a closed algebra for the local operators. It is the same reason why we only have partial results regarding applying circuit complexity to interacting quantum systems using different approximation schemes \cite{Bhattacharyya:2018bbv, Chowdhury:2023iwg, Haque:2024ldr}\footnote{Circuit complexity for circuit generated by symmetry generators for certain quantum field theories, \textit{e.g.} CFT, warped CFT and BMS$_3$ has been computed \cite{Caputa:2018kdj,Erdmenger:2020sup,Bhattacharyya:2022ren,Chagnet:2021uvi,Bhattacharyya:2023sjr}. But the connection to chaos has not been explored.}. This means systems which are potentially interesting from the point of view of chaos (in the sense of displaying truly chaotic dynamics as opposed to artefacts of having unstable saddle points) are precisely the ones hard to test using geometric complexity.

This is the gap that our paper tries to fill by applying a recently proposed algorithm to compute geometric complexity for interacting quantum mechanical systems \cite{Chowdhury:2023iwg, Chowdhury:2024ufv} to a coupled oscillator system. More concretely, we will be studying the Pullen-Edmonds model \cite{pullen1981comparison}:
\begin{equation}\label{Ham}
    \hat{H}=\frac{p_A^2}{2m}+\frac{p_B^2}{2m} + \frac{m\omega_m^2}{2}\left( x_A^2+x_B^2\right)+\lambda x_A ^2 x_B^2\, ,
\end{equation}
with $\lambda$ being positive, which is well-established as a truly chaotic system. In this case, the system is bounded by a positive potential and thus any  linear growth of complexity can be conjectured to be the diagnosis of quantum chaos rather than the result of any instability in the potential.  Our method, although providing only an upper bound on complexity, is an improvement on other approaches since it is not restricted to a small range of early-time behaviour \cite{Haque:2024ldr}, the latter being a regime in which it would be difficult to establish the chaotic nature of a system.

In Section 2, we start by reviewing this model using the RMT method to establish its chaotic antecedents. In Section 3, we give a detailed description of the geometric method that we will be using to compute the complexity for non-Gaussian systems. Then we compute the time dependence of complexity for our chaotic model. We are able to show that the geometric complexity reveals a linear growth for our chaotic system and is indeed a diagnosis to identify such behaviour. Thus, we use this model to benchmark the effective of circuit complexity to quantify chaos. We end by summarising our results and comparing this method to existing ones in the literature.

\section{Calculation from Random Matrix Theory} \label{sec2}
Random Matrix Theory (RMT) is a widely accepted diagnostic to probe quantum chaos. It states that the statistical behaviour of the spectrum of a quantum chaotic system can be described by that of a random matrix\footnote{There are some non-generic counterexamples \cite{PhysRevLett.69.1477}.} \cite{mehta2004random,ullmo2014introduction, Cotler:2016fpe}. A random matrix is a matrix with random variables as its entries, which all obey the same statistical properties. 

There are three random matrix ensembles that are relevant to the discussion depending on the symmetries of the quantum system. For an integrable system, according to Berry-Tabor Conjecture, its spectrum statistics are described by the Poisson Ensemble\footnote{Note that additional symmetries of the system will introduce commensurability of the spectra, which will lead to deviations from the Poisson PDF.}, with the Probability Density Function (PDF) of its level spacing, defined as the difference of two successive energy eigenvalues, given by:
\begin{equation}
    P_{\rm Poisson}(s)=e^{-s} 
\end{equation}
where $s$ is the dimensionless normalised level spacings where we have set the mean level spacing to $1$.

\begin{figure}[H]
    \centering
    \includegraphics[width=0.5\textwidth]{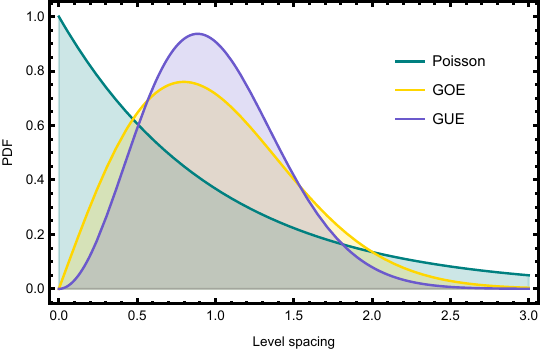}
    \caption{The PDFs corresponding to the three different ensembles.}
    \label{Fig1}
\end{figure}

For chaotic systems with time-reversal symmetry and those that do not have time-reversal symmetry, their level spacing statistics matches the Gaussian Orthogonal Ensemble (GOE) and the Gaussian Unitary Ensemble (GUE), respectively:
\begin{equation}
    P_{\rm GOE}(s)=\frac{\pi}{2}s e^{-\pi s^2 /4} ,\quad 
    P_{\rm GUE}(s)=\frac{32}{\pi^2}s^2 e^{-4 s^2 /\pi}  
\end{equation}
Their shapes are as shown in Fig.~\ref{Fig1}.

\begin{figure}[h]
    \centering
    \includegraphics[width=1\linewidth]{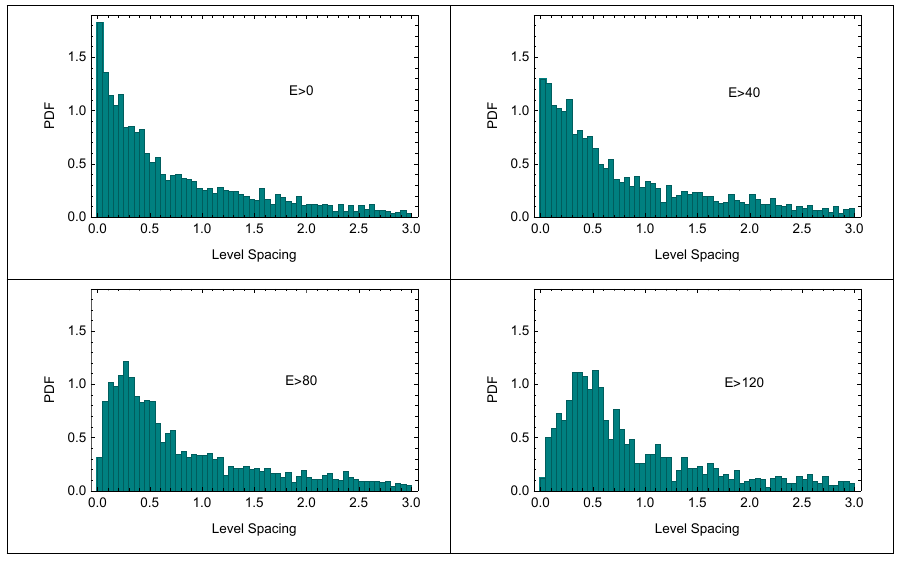}
    \caption{These are four PDFs of energy level spacings at different energy levels. As we can see, the system become more similar to GOE as the energy increases. }
    \label{fig:rmtcomparison}
\end{figure}

Our Hamiltonian \eqref{Ham} is clearly time-reversal symmetric, and thus we expect that the spectrum structure will be similar to that of GOE if it is indeed chaotic. For convenience of numerical computations, we set some of the parameters of the Hamiltonian \eqref{Ham} to unity to find
\begin{equation}
    H=\frac{p_{x}^{2}}{2}+\frac{p_{y}^{2}}{2}+\frac{x^2}{2}+\frac{y^2}{2}+\lambda x^{2} y^{2} \ ,
\end{equation}
with $\lambda > 0$. Note that this analysis to identify the chaotic nature of our Hamiltonian exists in the literature \cite{haller1984uncovering}, and we are simply reproducing those results to unify notation and for the ease of the reader. Classically, this model transfers from having quasi-periodicity to becoming chaotic as the energy increases. Correspondingly, there is degeneracy of energy eigenvalues at low energy level due to the symmetry $x\leftrightarrow y$, while at higher energy levels, the quartic potential introduces repulsion to the level spacing statistics. By setting $\lambda=0.1$, it can be shown through numerical plots (see Fig. \ref{fig:rmtcomparison}) that the system approaches GOE as we raise the energy cut off.

\begin{figure}[h]
    \centering
    \includegraphics[width=0.7\linewidth]{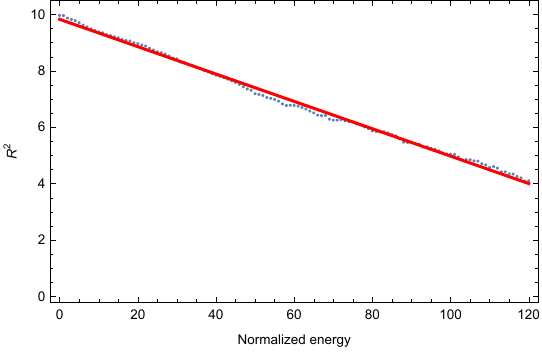}
    \caption{The x-axis is the normalised energy starting point, \textit{i.e.} we neglect all the energy eigenvalues below this value and only count the energy level spacing higher than that. The y-axis is the $\mathrm{R}^{2}$ as our fitting parameter. As we go to higher energies, the fitting number is smaller, meaning that the distribution of difference of spectrum is more likely to be GOE.}
    \label{fig:rmt-energy1}
\end{figure}

To capture the energy dependence of quantum chaos, we use the statistical quantity of residual squared $\mathrm{R}^2$ as our fitting parameter. Suppose we have data of a set of energy eigenvalue level spacings $\{a_{x_i}\}$, with $i$ running from 0 to n, and a fitting function $f(x)$, which for our case is the GOE density function, then the $\mathrm{R}^2$ is defined as:
\begin{equation}
    \mathrm{R}^2:=\sum_{i=0}^{n} \left[(a_{x_{i}}-f(x_{i}))^2\right]
\end{equation}
where we have ignored the normalising parameter since we are only interested in the comparison between different energy levels. The smaller the value of $\mathrm{R}^2$ is, the better the fitting function. We see a clear decreasing curve for $\mathrm{R}^{2}$ in Fig. \ref{fig:rmt-energy1}, implying that the system gets more chaotic as it goes to higher energy eigenvalues.

In conclusion, we recover the expected result that the Hamiltonian \eqref{Ham}, with a non-zero perturbation parameter $\lambda$, reveals a statistical behaviour similar to the Gaussian Orthogonal Ensemble, which we recognise as quantum chaos for a time-reversal symmetric system. In contrast, the $\lambda=0$ case is an obviously integrable system, with its level spacing values only being taken at $\delta E=0$ and $\delta E=1$, which is not described by GOE. Having proven that our system is a truly chaotic one for $\lambda >0$, that has been characterized by the RMT diagnostic, we next turn to compute the complexity of this system to see if we are able to recognize the chaotic dynamics in this system from this quantity.

\section{Chaos quantified by complexity}
\subsection{Review of geometric complexity}
Nielsen {\em et.\ al\/} \cite{Nielsen_2006,https://doi.org/10.48550/arxiv.quant-ph/0502070,https://doi.org/10.48550/arxiv.quant-ph/0701004} made new linkages between differential geometry and quantum complexity in a number of studies, proposing a change from the discrete explanation of gate complexity to continuous one. They noticed that the problem of determining the quantum complexity of a unitary operation is related to the problem of finding minimal-length geodesics in certain curved geometry. A geometrical definition of quantum complexity was first proposed as a tool to constrain the value of gate complexity; from there, it developed into a contender for a distinctive definition of quantum complexity. This is crucial to apply circuit complexity to quantum mechanical, and more generally, quantum field-theoretic systems.

In the geometrical framework, the length of the minimal geodesic on the unitary group manifold connecting the identity to the evolution operator $U$ is the complexity of this unitary operator. The minimal geodesic corresponds to the optimal circuit. An extension of the basic idea of the geometrical framework, which was based on unitaries acting on $n$-qubit systems, to a general unitary is initially straightforward, but it does lead to several mathematical subtleties  (see \cite{Chowdhury:2023iwg} for details). One begins by identifying the set of fundamental operators related to the target unitary operator and classifying them as ``easy'' or ``hard''. In order to define a geometry, one then considers a right-invariant metric ($G_{IJ}$), popularly known as the \textit{penalty factor matrix} that accurately penalizes the directions along the hard operators such that moving in their direction is discouraged for geodesics in the Lie group. The choice of the matrix $G_{IJ}$ is usually motivated by phenomenological considerations, inspired by difficulties of performing certain operations \cite{Brown:2019whu}. The choice of the metric leads to a notion of distance on the unitary space, which is given by
\begin{align}
\label{lineelement}
    ds^2 = \frac{1}{{\rm Tr}(\mathcal{O}_I\mathcal{O}^{\dagger}_I) {\rm Tr}(\mathcal{O}_J\mathcal{O}^{\dagger}_J)}\bigg[G_{IJ} {\rm Tr}[i U^{-1}\mathcal{O}_I^{\dagger}dU] {\rm Tr}[i U^{-1}\mathcal{O}_J^{\dagger}dU]\bigg],
\end{align}
where the $\mathcal{O}_I$ represents the generators of the unitary group and $U$ plays the role 
of a point on the manifold. The trace ${\rm Tr}$ is taken in a matrix representation of the generators\footnote{In \cite{Bhattacharyya:2018bbv}, a generalization to studying interacting QFTs was based on using finite-dimensional matrices for these opeartors while in \cite{Haque:2024ldr} a truncation was sought based on a power-series expansion in time.}. For geodesics, only the right-invariance of the line element matters, but not the specific form on the entire group. To see this in more detail, recall that an efficient way of determining geodesics on Lie groups equipped with a right invariant metric is by using the so-called Euler-Arnold equation \cite{AIF_1966__16_1_319_0}\footnote{An English translation of the original French article can be found here: \hyperlink{https://terrytao.wordpress.com/2010/06/07/the-euler-arnold-equation/}{https://terrytao.wordpress.com/2010/06/07/the-euler-arnold-equation/}.}, which has been extensively 
used recently to compute the geodesics on unitary manifolds \cite{Chowdhury:2023iwg, Chowdhury:2024ufv,Balasubramanian:2019wgd,Balasubramanian:2021mxo, 
Flory:2020dja, Haque:2024ldr}. The Euler-Arnold equations uses the structure constants of the Lie algebra corresponding to the group. No other information regarding the group or its form is necessary while solving it. The necessity of the right invariance of the line element can be understood from its derivation provided in the appendix of \cite{Chowdhury:2023iwg}. But more importantly the Euler-Arnold equation gives geodesic flow on a manifold equipped with a right invariant metric.

They are given by
\begin{equation}
\label{eqn:eulerarnoldrev}
	G_{IJ}\frac{dV^{J}(s)}{ds}= f_{IJ}^{K} V^{J}(s) G_{KL}V^{L}(s)\,,
\end{equation}
where $f_{IJ}^{K}$ are the structure constants of the Lie algebra, defined by
\begin{equation}
\label{structure}
	[\mathcal{O}_{I},\mathcal{O}_{J}]= i f_{IJ}^{K} \mathcal{O}_{K}\,.
\end{equation}
The components $V^I(s)$ represent the tangent vector (or the velocity) at each point along the geodesic, defined by:
\begin{align}
\label{differentialU}
	\frac{dU(s)}{ds}=-i V^I(s)\mathcal{O}_I U(s)\,.
\end{align}
Given a solution $V^I(s)$, a further integration of (\ref{differentialU}) results in the path (or trajectory) in the group, guided by the velocity vector $V^I(s)$. Generically, this solution can be written as  the path-ordered exponential
\begin{equation}
\label{sdependentunitary}
	 U(s)=\mathcal{P}\exp\bigg(-i\int_{0}^{s}ds'~V^I(s')\mathcal{O}_I\bigg)\,,
\end{equation}
on which we impose the boundary conditions: 
\begin{equation}
\label{boundary}
	U(s=0)=\mathbb{I} ~~~~{\rm and}~~~ U(s=1)=U_{\rm target}\,.
\end{equation}
where $U_{\rm target}$ is some target unitary whose complexity we wish to study. 

In general, equation (\ref{eqn:eulerarnoldrev}) defines a family of geodesics $\{V^I(s)\}$ on the unitary space. The boundary condition $U(s=1)=U_{\rm target}$ filters out those geodesics that can realize the target unitary operator by fixing the magnitude of the tangent vector $V^I$ at $s=0$ (at the identity operator). There could be more than one value of the $v_i$'s for which the point of the target unitary is reached. However, the smallest value is to be considered as we are looking for the shortest geodesic.
\begin{equation}
\label{eqn:complexityexpression}
	C[U_{\rm target}]= \stackunder{{\rm min}}{\{$V^I(s)$\}}\int_{0}^{1}ds\sqrt{G_{IJ}V^{I}(s)V^{J}(s)}\,,
\end{equation}
where the minimization is over all geodesics $\{V^I(s)\}$ from the identity to $U_{\rm target}$. 

\subsection{Geometric complexity of the considered model} 
Now we turn our attention to the following Hamiltonian:
\begin{equation}
    \hat{H}=\frac{p_A^2}{2m}+\frac{p_B^2}{2m} + \frac{m\omega_m^2}{2}\left( x_A^2+x_B^2\right)+\lambda x_A ^2 x_B^2.
\end{equation}
We  choose $m=\omega_m=1$ to match with our analyses from the previous section, and we relegate the more general case to the Appendix~\ref{App}. This allows us to write the Hamiltonian in the following form:
\begin{align}
    H= \frac{p_A^2}{2}+\frac{x_A^2}{2}+ \frac{p_B^2}{2}+\frac{x_B^2}{2}+\lambda x_A^2x_B^2.
\end{align}
We can consider the following generators:
\begin{align}
     M_1 &= \frac{p_A^2}{2}+\frac{x_A^2}{2}, ~~~ M_2= \frac{p_B^2}{2 }+\frac{ x_B^2}{2}, \\
    M_3 &= \frac{p_A^2}{2}-\frac{x_A^2}{2}, ~~~ M_4= \frac{p_B^2}{2}-\frac{x_B^2}{2}, \\
    M_5 &= \frac{1}{2}(x_Ap_A+p_Ax_A), ~~~ M_6 = \frac{1}{2}(x_Bp_B+p_Bx_B), \\
    M_7 &= x_A^4, ~~~ M_8= x_B^4, ~~~ M_9= x_A^2x_B^2.
\end{align}
In terms of these generators, the Hamiltonian can be written as: 
\begin{align}
    H= \omega_m M_1+\omega_m M_2 + \lambda M_9.
\end{align}
Therefore, our target unitary operator is: 
\begin{align}
\label{targetunitary}
    U_{\rm target} = \exp\bigg(-i\bigg(\omega_m M_1+\omega_m M_2+\lambda M_9 \bigg)t\bigg).
\end{align}
Now if we take the commutators, we find: 
\begin{align}
    [M_1,M_3] &= 2i M_5, \\
    [M_1,M_5] &= -2i M_3, \\
    [M_3,M_5] &= -2i M_1 ,\\
    [M_1,M_2] &= 0, \\
    [M_1,M_4] &=0 ,\\
    [M_1, M_6] &= 0, \\
    [M_2,M_4] &= 2i M_6, \\
    [M_2,M_6] &= -2i M_4, \\
    [M_4,M_6] &= -2i M_2, \\
    [M_1,M_2] &= 0, \\
    [M_1,M_4] &=0, \\
    [M_1, M_6] &= 0, \\
    [M_1,M_7] &= -i(x_A^3p_A+x_A^2p_Ax_A+x_Ap_Ax_A^2+p_Ax_A^3) = -i M_{10}, \\
    [M_1,M_8] &= 0, \\
    [M_1,M_9] &= -i (x_Ap_Ax_B^2+p_Ax_Ax_B^2) = -i M_{11}.
\end{align}
This is where we have implicitly made our first approximation. For interacting systems such as ours, the algebra of the operators do not naturally close and is an infinite-dimensional one, as expected. However, we have only kept terms to $\mathcal{O}(4)$ order. We are already implicitly forcing our geodesics to be in the lower-dimensional manifold of the full infinitely-dimensional space by truncating this algebra. In short, this (and neglecting higher-order terms in the Dyson series, as will be explained shortly) is why we can only claim to find an upper bound on complexity since there might always be a shorter geodesic available in the higher-dimensional direction that we have left out (see \cite{Chowdhury:2023iwg} for more details.)

Using the knowledge obtained from the commutation relation of the operators, we can proceed to solve the Euler-Arnold equations \eqref{eqn:eulerarnoldrev} which, in our case, can be written more explicitly as:
\begin{align}
    G_{11}\frac{dV^1}{ds} \approx & f_{13}^5 V^3G_{55}V^5 + f_{15}^3 V^5G_{33}V^3+ f_{17}^{10}V^7G_{1010}V^{10}+f_{19}^{11}V^9G_{1111}V^{11}, \\
    G_{22}\frac{dV^2}{ds} \approx & f_{24}^6 V^4 G_{66}V^6+ f_{26}^{4}V^6G_{44}V^4 + f_{28}^{12}V^8G_{1212}V^{12}+ f_{29}^{13}V^{9}G_{1313}V^{13},  \\
    G_{33}\frac{dV^3}{ds} \approx & f_{31}^5 V^1 G_{55}V^5+ f_{35}^1 V^5 G_{11}V^1 + f_{37}^{10}V^7G_{1010}V^{10} + f_{39}^{11}V^9G_{1111}V^{11}, \\
    G_{44}\frac{dV^4}{ds} \approx & f_{42}^6 V^2 G_{66}V^6+ f_{46}^2 V^6 G_{22}V^2 + f_{48}^{12}V^8G_{1212}V^{12}+ f_{49}^{13}V^9G_{1313}V^{13}, \\
    G_{55}\frac{dV^5}{ds} \approx & f_{51}^3 V^1 G_{33}V^3 + f_{53}^1 V^3 G_{11}V^1 + f_{57}^7 V^7 G_{77}V^7 + f_{59}^9 V^9 G_{99}V^9, \\
    G_{66}\frac{dV^6}{ds} \approx & f_{62}^4 V^2 G_{44}V^4 + f_{64}^2 V^4G_{22}V^2 + f_{68}^8 V^8 G_{88}V^8 + f_{69}^9 V^9G_{99}V^9, \\
    G_{77}\frac{dV^7}{ds} \approx & f_{71}^{10}V^{1}G_{1010}V^{10}+ f_{73}^{10}V^3G_{1010}V^{10}+ f_{75}^7 V^5G_{77}V^7, \\
    G_{88}\frac{dV^8}{ds} \approx & f_{82}^{12}V^2G_{1212}V^{12}+ f_{84}^{12}V^4G_{1212}V^{12}+ f_{86}^8 V^6G_{88}V^8, \\
    G_{99}\frac{dV^9}{ds} \approx & f_{91}^{11}V^1G_{1111}V^{11} + f_{92}^{13}V^2G_{1313}V^{13} + f_{93}^{11}V^3G_{1111}V^{11} \\ & + f_{95}^9 V^5 G_{99}V^9 + f_{96}^9 V^6G_{99}V^9 .
\end{align}
This is where further approximations are required. In principle, we should consider all the quartic operators but we would truncate up to those terms that appear in the target unitary operator. While setting up the Euler-Arnold equation, we have used the ``$\approx$" sign instead of the ``=" sign as some of the terms have been neglected. The following arguments justify why one should be able to do that:
\begin{itemize}
    \item \textbf{Step 1}: The penalties associated with all the quadratic operators are identical i.e $G_{11}=G_{22}=G_{33}=G_{44}=G_{55}=G_{66}=p$.
    \item \textbf{Step 2}: The penalties associated with all the quartic operators are equal ($q$) and large ($q>>p)$ such that the component of the velocity vector $V^I(s)$ along those directions are negligible \cite{Bhattacharyya:2018bbv}. The penalties of operators of order higher than quartic order also have prohibitive penalties such that the corresponding $V^I$'s $\rightarrow$ 0.
\end{itemize}
These go in line with our primary goal of preferring the lower dimensional manifold (corresponding to the `free' system described by  the quadratic operators) and penalizing geodesics that go in the higher-dimensional directions. Once again, the basis of this is that we are interested in a `perturbative' notion of geometric complexity that provides a sensible and rigorous upper bound for us.

The above two approximations considerably simplify the Euler-Arnold equations. Let us explain in a little more detail. Consider the equation corresponding to the first generator i.e $\frac{dV^1}{ds}$. It has terms like $f_{17}^{10} V^7 G_{1010}V^{10}$, $f_{19}^{11}V^9 G_{1111}V^{11}$. Owing to the large penalties associated with $M_7$, $M_8$, $M_9$ and $M_{10}$ the corresponding components $V^7$, $V^8$, $V^9$, $V^{10}$ would be negligible. Hence, the contribution of the product of such terms would be extremely small and hence can be neglected. Using the above argument we can rewrite the above equations as: 
\begin{align}
     \frac{dV^1}{ds} \approx & f_{13}^5 V^3V^5 + f_{15}^3 V^5V^3, \\
    \frac{dV^2}{ds} \approx & f_{24}^6 V^4V^6+ f_{26}^{4}V^6V^4,  \\
    \frac{dV^3}{ds} \approx & f_{31}^5 V^1V^5+ f_{35}^1 V^5V^1, \\
    \frac{dV^4}{ds} \approx & f_{42}^6 V^2V^6+ f_{46}^2 V^6V^2, \\
    \frac{dV^5}{ds} \approx & f_{51}^3 V^1V^3 + f_{53}^1 V^3V^1, \\
    \frac{dV^6}{ds} \approx & f_{62}^4 V^2V^4 + f_{64}^2 V^4V^2, \\
    \frac{dV^7}{ds} \approx & f_{71}^{10}V^{1}V^{10}+ f_{73}^{10}V^3V^{10}+ f_{75}^7 V^5V^7, \\
    \frac{dV^8}{ds} \approx & f_{82}^{12}V^2V^{12}+ f_{84}^{12}V^4V^{12} + f_{86}V^6V^8,\\
    \frac{dV^9}{ds} \approx & f_{91}^{11}V^1V^{11} + f_{92}^{13}V^2V^{13} + f_{93}^{11}V^3V^{11}+ f_{95}^9 V^5V^9 + f_{96}^9 V^6V^9\,.
\end{align}
The equations $\frac{dV^7}{ds}$, $\frac{dV^8}{ds}$ and $\frac{dV^9}{ds}$ could be immediately simplified. Since the penalties associated with the quartic operators $M_7$, $M_8$,$M_9$, $M_{10}$, $M_{11}$ and $M_{12}$, are much larger compared to the quadratic operators ($q>>p$), the components $V^7$, $V^8$, $V^9$, $V^{10}$, $V^{11}$ and $V^{12}$ are extremely small. Therefore, the products $V^IV^J$ where $1 \leq I\leq 6$ and $7\leq I \leq 13$ would be negligible as $V^I$'s for $1 \leq  I \leq 6$ though finite are not divergent quantities. If we take into consideration this argument, then the equations $\frac{dV^7}{ds}$, $\frac{dV^8}{ds}$ and $\frac{dV^9}{ds}$ can be greatly simplified and can, in fact, be equal to zero.

With these approximations at hand, we can finally express the Euler-Arnold equations as:
\begin{align}
    \frac{dV^1}{ds} &\approx 0, \\
    \frac{dV^2}{ds} &\approx 0, \\
    \frac{dV^3}{ds} &\approx -4 V^1 V^5, \\
    \frac{dV^4}{ds} &\approx -4 V^2V^6, \\
    \frac{dV^5}{ds} &\approx 4 V^1V^3, \\
    \frac{dV^6}{ds} &\approx 4V^2V^4, \\
    \frac{dV^7}{ds} &\approx 0, ~~~ \frac{dV^8}{ds} \approx 0, ~~~~ \frac{dV^9}{ds} \approx 0\,.
\end{align}
The solutions of these equations can be written as: 
\begin{align}
    V^1(s) &= v_1, ~~ V^2(s)= v_2, \\
    V^3(s) &= v_3 \cos (4 s v_1)-v_5 \sin (4 s v_1), \\
    V^4(s) &= v_4 \cos (4 s v_2)-v_6 \sin (4 s v_2) ,\\
    V^5(s) &= v_5 \cos (4 s v_1)+v_3 \sin (4 s v_1), \\
    V^6(s) &= v_6 \cos (4 s v_2)+v_4\sin (4 s v_2), \\
    V^7(s) &= v_7, ~~~V^8(s)= v_8, ~~~ V^9(s)= v_9\,.
\end{align}

The complexity of the target unitary operator in terms of the $v_i$'s is thus given by: 
\begin{align}
C[U_{\rm target}]= \sqrt{p(v_1^2+v_2^2+v_3^2+v_4^2+v_5^2+v_6^2)+q(v_7^2+v_8^2+v_9^2)}\,.
\end{align}
We want to know the geodesic for fixed boundary conditions 
$U(s=0)= \mathbb{I}$ and $U(s)= U_{\rm target}$ to fix the $v_i$'s. 
The unitary along the geodesic path from the identity with a specific 
tangent vector $V(s)$ is given by the path-ordered exponential:
\begin{align}
    U(s)= \mathcal{P} \exp\bigg(-i\int_0^s V^I(s') M_I ds'\bigg)\,,
\end{align}
which is a solution to the equation: 
\begin{align}
\label{differentialU1}
    \frac{dU(s)}{ds}=-i V^I(s)M_I U(s)\,. 
\end{align}
Solving $U(s)$ would require dealing with the path ordering, 
which is a notoriously difficult problem and is usually solved using an iterative approach, which is expressed through the \textit{Dyson expansion}. However, to keep the computation tractable, we ignore the path ordering. Neglecting the path ordering does not give us the precise value of the complexity because we will no longer obtain the shortest trajectory but some other trajectory whose length would be more than the desired path. Hence, we will only be able to interpret our result as an \textit{upper bound}\footnote{An interpretation on the upper bound of complexity was also provided in \cite{Craps:2022ese,Craps:2023rur}.} on complexity. Therefore, we approximate the unitary operator as:
\begin{align} \label{Un}
    U(s) \approx \exp\bigg(-i\int_0^s V^I(s')M_I ds'\bigg).
\end{align}
With the obtained solutions of the Euler-Arnold equations, we can write the above expression as: 
\begin{align}
\nonumber
    U(s) &\approx \exp\bigg(-i\bigg(v_1 s M_1 + v_2 s M_2 + \bigg(\frac{v_3 \sin (4 s v_1)-2 v_5 \sin ^2(2 s v_1)}{4 v_1}\bigg)M_3 \\ \nonumber & ~~~ + \bigg(\frac{v_4 \sin (4 s v_2)-2 v_6 \sin ^2(2 s v_2)}{4 v_2}\bigg)M_4+ \bigg(\frac{2 v_3 \sin ^2(2 s v_1)+v_5 \sin (4 s v_1)}{4 v_1}\bigg)M_5 \\ & ~~~~~~~ +\bigg(\frac{2 v_4 \sin ^2(2 s v_2)+v_6 \sin (4 s v_2)}{4 v_2}\bigg)M_6 +v_7 s M_7+ v_8 s M_8+ v_9 s M_9+...\bigg)\bigg)\,.
\end{align}

The above equation at $s=1$ should reach $U_{\rm target}$, \textit{i.e.}
\begin{align}
\label{targetunitary}
   U(s=1)= U_{\rm target} = \exp\bigg(-i\bigg(M_1+ M_2+\lambda M_9 \bigg)t\bigg)\,,
\end{align}
which gives us the following values of the integration constants obtained from the Euler-Arnold equations:
\begin{align}
    v_1= t, ~~~ v_2= t, ~~~ v_9= \lambda t\,,
\end{align}
which gives us the complexity of $U_{\rm target}$ as follows:
\begin{align}
    C[U_{\rm target}]= \sqrt{p(v_1^2+v_2^2)+q \lambda ^2 t^2}\,.
\end{align}

One must not forget that there is a periodicity associated with $v_1$ and $v_2$. The generators $M_1$ and $M_2$ that correspond to the Hamiltonian of the individual oscillators has spectrum $n_1+1/2$ and $n_2+1/2$, with integer $n_1$ and $n_2$. The operators $M_1+M_2$ and $M_1-M_2$ thus has an integer spectrum. Therefore, on embedding the finite dimensional Lie group used here in the infinite dimensional Hilbert space of quantum mechanics, the generators $M_1+M_2$ and $M_1-M_2$ should exponentiate to an operator with period $2\pi$. To exploit the condition of periodicity, we note that:
\begin{align}
    v_1 M_1+v_2 M_2 = \frac{1}{2}\bigg((v_1+v_2)(M_1+M_2)+(v_1-v_2)(M_1-M_2)\bigg).
\end{align}
From the above equation, it can be seen that a period of $4\pi$ for $v_1+\pm v_2$ is necessary to generate of period of $2\pi$ for $M_1\pm M_2$.
Therefore, to invoke the periodicity argument let's rewrite the complexity expression as:
\begin{align}
    C[U_{\rm target}] = \sqrt{\frac{p}{2}\bigg((v_1+v_2)^2+(v_1-v_2)^2\bigg)+q \lambda^2 t^2}.
\end{align}
which can be further simplified to: 
\begin{align}
    C[U_{\rm target}]= \sqrt{\frac{p}{2}(v_1+v_2)^2+q\lambda^2t^2},
\end{align}
as $v_1-v_2=0$ in this case.

Writing the complexity expression explicitly in terms of $v_1$ and $v_2$ allows us to implement the periodicity condition associated with the generators $M_1$ and $M_2$ in the complexity. To study the behavior of the complexity upper bound we need to fix the values of the penalty factors $p$ and $q$. Although, the choice of penalty factors may seem arbitary, it is not the case here. It has an intricate relation with the interaction parameter $\lambda$. In our computation, we resorted to perturbation theory but the argument was based on the penalty factors $p$ and $q$. Thus there must be some relation between these penalty factors and $\lambda$. The parameter $q$ was associated with the operators of higher order and indicated a higher penalty, hence it should be inversely related to the parameter $\lambda$ (as the interaction strength is small for our case), whereas $p$, being the lower penalty term should be linked with $\lambda$ directly. Therefore,
\begin{align}
\label{choicepenalty}
    q \sim \frac{1}{\lambda}, ~~~~~ p \sim \lambda\,.
\end{align}
It must be noted that when $\lambda=0$, the higher order operators do not play any further role in the analysis. The Lie algebra of operators closes at the quadratic level and hence all the operators should have equal penalties. In such scenario, the choice of having different penalties is not necessary any longer. Consequently, \eqref{choicepenalty} is valid only when $\lambda >0$.

\begin{figure}[H]
    \centering
    \includegraphics[scale=0.6]{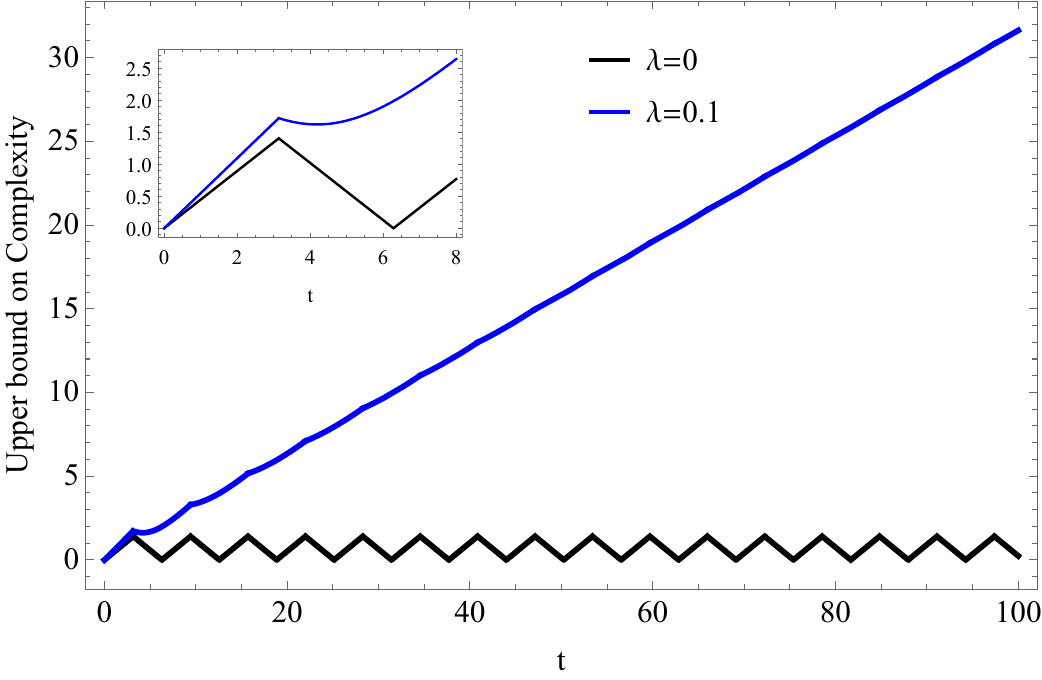}
    \caption{Behavior of the complexity. $q=10$, $p=0.1$.}
    \label{plot1}
\end{figure}

From  the Fig.~(\ref{plot1}), we see that when the interaction term is switched off, \textit{i.e.} when $\lambda=0$, the complexity exhibits the oscillatory behaviour as expected for a harmonic system. However, as soon as we switch on the interaction term, the complexity starts exhibiting a linearly growing behaviour which we identify as the quantifier of chaotic dynamics.

\begin{figure}[H]
    \centering
    \includegraphics[scale=0.6]{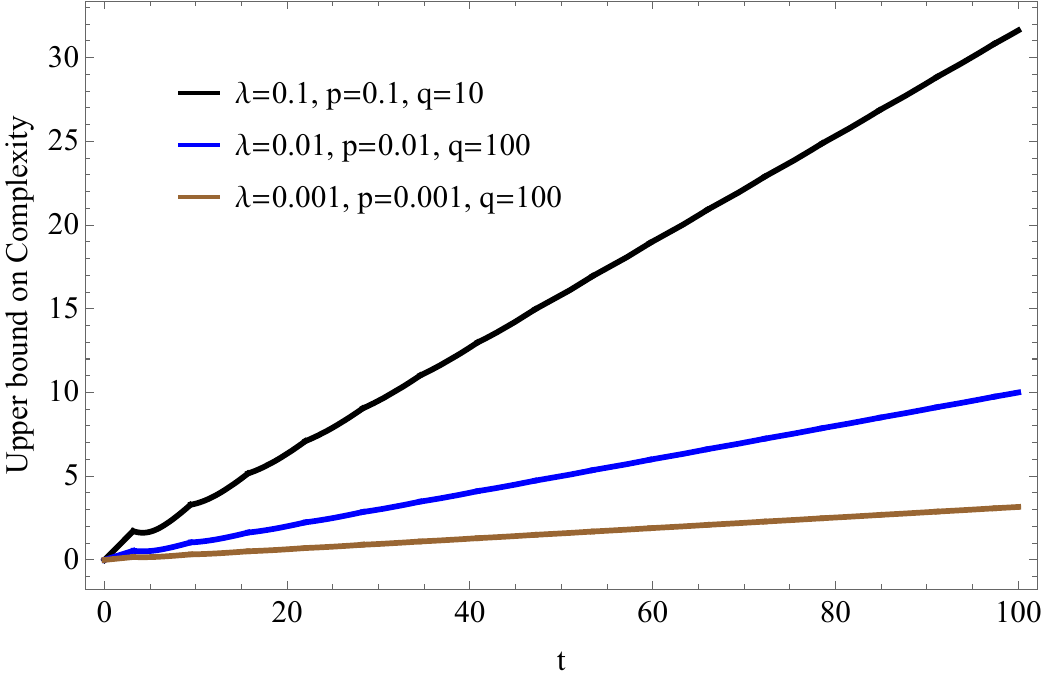}
    \caption{Behavior of complexity for different values of $\lambda$}
    \label{plot2}
\end{figure}

The robustness of the linearly growing behaviour of complexity is demonstrated in figure \ref{plot2} by choosing different values of $\lambda$.

\section{Conclusions and Discussion}
In this paper, we have initiated a study of the geometric complexity of the time evolution operator for a non-Gaussian system. We consider a non-Gaussian quantum mechanical model, namely the Pullen-Edmonds model. Then, we use the geometric method pioneered by Nielsen et al. \cite{Nielsen_2006} to study the circuit complexity of the time evolution operator. However, algebra for the generators for our case does not close naturally. For our analysis, we have truncated it by restricting ourselves to the quartic generators made out of the basic operators $x$ and $p$. Subsequently, we approximate the Euler-Arnold equation needed for computing the complexity by choosing the metric $G_{IJ}$ such that all the quadratic operators are weighted by the same penalty factor $p$ and all the quartic operators are weighted by these same penalty factors $q$ and $q\gg p.$ In this way, we can neglect the contributions to the velocity (tangent) vectors along those directions. This is a reasonable assumption because the more complicated the operators, the cost of making the operators should be higher from a simulation perspective. This enables us to compute the complexity without expanding in time, as done in previous studies. Furthermore, since we are focusing on the upper bound of complexity, we keep only the leading terms in the Dyson series of the path-ordered exponential of the optimized unitary \eqref{Un}. Finally, we see that, as soon as we turn on the quartic coupling in the Hamiltonian, the behaviour of complexity (after making appropriate identification of the penalty factors with the interaction strength \eqref{choicepenalty}) changes drastically from that of the free theory. It starts to grow linearly compared to the case where the underlying theory is free. This we can identify as the quantifier of chaotic dynamics and is consistent with the observations from RMT, where even turning on a small $\lambda$ can make this theory chaotic, as evident from the nearest-neighbour energy eigenvalue distribution that we discussed in Section~\eqref{sec2}. Finally, we would like to emphasize that, unlike \cite{Haque:2024ldr}, we do not expand the unitary operator for which we are computing the complexity (and the tangent vectors) in a power series in physical time. This makes the early time behaviour\footnote{This expansion describes the early time behaviour  since the series expansion in valid within a small time domain.} of the complexity independent of the algebra of the generators as claimed in \cite{Haque:2024ldr}. On the other hand, our approximation is a different one that ignores the path ordering of the exponential, for the given path along which we compute the complexity, leading to an upper bound on the complexity. Nevertheless, we have demonstrated within a set of simplifying assumptions how the circuit complexity can clearly distinguish between a chaotic and non-chaotic regime (algebra of the generators changes between these two regimes).

Our analysis paves the way forward for various future investigations. We have only shown that geometric complexity is a good complementary probe for quantum chaos. However, it remains to be seen if complexity can act as a quantifier of quantum chaos in systems where other measures fail for some reason or another. This will be explored in future work. Moreover, in this paper, we have mainly focused on a quantum mechanical system. It will be interesting to extend our study to interacting quantum field theories. The simplest example will be an interacting scalar field theory and study of this geometric complexity for the time evolution operator. Furthermore, there are recent proposals for a non-Gaussian cMERA (continuous multi-scale entanglement renormalization ansatz) tensor network for interacting quantum field theory \cite{Fernandez-Melgarejo:2020fzw}. It will be good to extend our computation for such cMERA circuits, which will help us to connect with the simulation of certain quantum field theories. Last, but not least, it will also be interesting to extend our analysis for non-Gaussian models of cosmology, thereby extending the studies of \cite{Bhattacharyya:2020rpy,Bhattacharyya:2020kgu} made in the context of states to the time evolution operator.

\acknowledgments
We thank Bret Underwood for comments on an earlier version of this draft. AB is supported by the Core Research Grant (CRG/2023/001120) and Mathematical Research Impact Centric Support Grant (MTR/2021/ 000490) by the Department of Science and Technology Science and Engineering Research Board (India). SB is supported in part by the Higgs Fellowship and by the STFC Consolidated Grant ``Particle Physics at the Higgs Centre''. SC would like to thank the doctoral school of Jagiellonian University for providing a fellowship during the course of the work. SC was supported by ``Research support module" as part of the ``Excellence Initiative – Research University" program at the Jagiellonian University in Kraków for this project. SC also acknowledges the hospitality of the Higgs Centre for Theoretical Physics at the University of Edinburgh during his visit where part of the work was conducted.  XL is supported in part by the Program of China Scholarship Council (Grant No. 202208170014).

\appendix

\section{Complexity derived from the general Hamiltonian}\label{App}
In this Appendix, we restore the parameters set to $1$ in the main body of the paper to compute the general expression for circuit complexity of this model. We have the following Hamiltonian:
\begin{equation}
    \hat{H}=\frac{p_A^2}{2m}+\frac{p_B^2}{2m} + \frac{m\omega_m^2}{2}\left( x_A^2+x_B^2\right)+\lambda x_A ^2 x_B^2.
\end{equation}

 Let us consider the following defined generators. 
\begin{align}
    M_1 &= \frac{p_A^2}{2m \omega_m^2}+\frac{m \omega_m^2 x_A^2}{2}, ~~~ M_2= \frac{p_B^2}{2 m\omega_m^2}+\frac{m \omega_m^2 x_B^2}{2}, \\
    M_3 &= \frac{p_A^2}{2m \omega_m^2}-\frac{m \omega_m^2 x_A^2}{2}, ~~~ M_4= \frac{p_B^2}{2 m\omega_m^2}-\frac{m \omega_m^2 x_B^2}{2}, \\
    M_5 &= \frac{1}{2}(x_Ap_A+p_Ax_A), ~~~ M_6 = \frac{1}{2}(x_Bp_B+p_Bx_B), \\
    M_7 &= x_A^4, ~~~ M_8= x_B^4, ~~~ M_9= x_A^2x_B^2.
\end{align}
In terms of these generators, the Hamiltonian can be written as: 
\begin{align}
    H= \frac{1}{2}\bigg(\omega_m^2 (M_1+M_3)+\omega_m^2 (M_2+M_4)+ (M_1-M_3)+ (M_2-M_4)+ \lambda M_9 \bigg).
\end{align}
Therefore, our target unitary operator is: 
\begin{align}
\label{targetunitary}
    U_{\rm target} = \exp\bigg(-i\bigg(\frac{1+\omega_m^2}{2}M_1+\frac{1+\omega_m^2}{2}M_2+\frac{\omega_m^2-1}{2}M_3+\frac{\omega_m^2-1}{2}M_4+\frac{\lambda}{2}M_9 \bigg)t\bigg).
\end{align}
Now, if we take the commutator we see: 
\begin{align}
    [M_1,M_3] &= 2i M_5, \\
    [M_1,M_5] &= -2i M_3, \\
    [M_3,M_5] &= -2i M_1, \\
    [M_1,M_2] &= 0, \\
    [M_1,M_4] &=0, \\
    [M_1, M_6] &= 0, \\
    [M_2,M_4] &= 2i M_6, \\
    [M_2,M_6] &= -2i M_4, \\
    [M_4,M_6] &= -2i M_2, \\
    [M_1,M_2] &= 0, \\
    [M_1,M_4] &=0, \\
    [M_1, M_6] &= 0, \\
    [M_1,M_7] &= -\frac{i}{\omega_m^2}(x_A^3p_A+x_A^2p_Ax_A+x_Ap_Ax_A^2+p_Ax_A^3) = -i M_{10}, \\
    [M_1,M_8] &= 0, \\
    [M_1,M_9] &= -\frac{i}{\omega_m^2} (x_Ap_Ax_B^2+p_Ax_Ax_B^2) = -i M_{11}.
\end{align}

With the obtained solutions of the Euler-Arnold equations, we can write:
\begin{align}
\nonumber
    U(s) &\approx \exp\bigg(-i\bigg(v_1 s M_1 + v_2 s M_2 + \bigg(\frac{v_3 \sin (4 s v_1)-2 v_5 \sin ^2(2 s v_1)}{4 v_1}\bigg)M_3 \\ \nonumber & ~~~ + \bigg(\frac{v_4 \sin (4 s v_2)-2 v_6 \sin ^2(2 s v_2)}{4 v_2}\bigg)M_4+ \bigg(\frac{2 v_3 \sin ^2(2 s v_1)+v_5 \sin (4 s v_1)}{4 v_1}\bigg)M_5 \\ & ~~~~~~~ +\bigg(\frac{2 v_4 \sin ^2(2 s v_2)+v_6 \sin (4 s v_2)}{4 v_2}\bigg)M_6 +v_7 s M_7+ v_8 s M_8+ v_9 s M_9+...\bigg)\bigg).
\end{align}

The above equation at $s=1$ should give $U_{\rm target}$ i.e
\begin{align}
    U(s=1)= \exp\bigg(-i\bigg(\frac{1+\omega_m^2}{2}M_1+\frac{1+\omega_m^2}{2}M_2+\frac{\omega_m^2-1}{2}M_3+\frac{\omega_m^2-1}{2}M_4+\frac{\lambda}{2}M_9\bigg)t\bigg).
\end{align}

Comparing the coefficients of the generators $M_1$ and $M_2$, in the exponential we get 
\begin{align}
    v_1= \bigg(\frac{1+\omega_m^2}{2}\bigg)t, ~~~ v_2 = \bigg(\frac{1+\omega_m^2}{2}\bigg)t.
\end{align}

 As discussed earlier, the periodicity associated with the generators $M_1+M_2$ and $M_1-M_2$ should be taken into account.

 Therefore, it is much more suitable to express the other integration constants $v_i's$ in terms of $v_1$ and $v_2$ as follows: 
 \begin{align}
     v_3 & = v_1 \left(\omega_m^2-1\right) \cot (2 v_1), \\
     v_4 &= v_2 \left(\omega_m^2-1\right) \cot (2 v_2), \\
     v_5 &= v_1-v_1 \omega_m^2, \\
     v_6 &= v_2-v_2 \omega_m^2.
 \end{align}

In terms of these $v_i$'s the complexity expression can be written as: 
\begin{align}
    & C[U_{\rm target}] =  \nonumber \\
    & \sqrt{p \left(\left(\omega_m^4-2 \omega_m^2+2\right) \left(v_1^2+v_2^2\right)+v_1^2 \left(\omega_m^2-1\right)^2 \cot ^2(2 v_1)+v_2^2 \left(\omega_m^2-1\right)^2 \cot ^2(2 v_2)\right)+q \frac{\lambda^2 t^2}{4}}.
\end{align}






\bibliography{biblio}

\providecommand{\href}[2]{#2}\begingroup\raggedright\begin{thebibliography}{10}

\bibitem{DAlessio:2015qtq}
L.~D'Alessio, Y.~Kafri, A.~Polkovnikov, and M.~Rigol, ``{From quantum chaos and eigenstate thermalization to statistical mechanics and thermodynamics},'' \href{http://dx.doi.org/10.1080/00018732.2016.1198134}{{\em Adv. Phys.} {\bfseries 65} no.~3, (2016) 239--362}, \href{http://arxiv.org/abs/1509.06411}{{\ttfamily arXiv:1509.06411 [cond-mat.stat-mech]}}.

\bibitem{Shenker:2013pqa}
S.~H. Shenker and D.~Stanford, ``{Black holes and the butterfly effect},'' \href{http://dx.doi.org/10.1007/JHEP03(2014)067}{{\em JHEP} {\bfseries 03} (2014) 067}, \href{http://arxiv.org/abs/1306.0622}{{\ttfamily arXiv:1306.0622 [hep-th]}}.

\bibitem{Shenker:2013yza}
S.~H. Shenker and D.~Stanford, ``{Multiple Shocks},'' \href{http://dx.doi.org/10.1007/JHEP12(2014)046}{{\em JHEP} {\bfseries 12} (2014) 046}, \href{http://arxiv.org/abs/1312.3296}{{\ttfamily arXiv:1312.3296 [hep-th]}}.

\bibitem{Shenker:2014cwa}
S.~H. Shenker and D.~Stanford, ``{Stringy effects in scrambling},'' \href{http://dx.doi.org/10.1007/JHEP05(2015)132}{{\em JHEP} {\bfseries 05} (2015) 132}, \href{http://arxiv.org/abs/1412.6087}{{\ttfamily arXiv:1412.6087 [hep-th]}}.

\bibitem{Roberts:2014isa}
D.~A. Roberts, D.~Stanford, and L.~Susskind, ``{Localized shocks},'' \href{http://dx.doi.org/10.1007/JHEP03(2015)051}{{\em JHEP} {\bfseries 03} (2015) 051}, \href{http://arxiv.org/abs/1409.8180}{{\ttfamily arXiv:1409.8180 [hep-th]}}.

\bibitem{Maldacena:2015waa}
J.~Maldacena, S.~H. Shenker, and D.~Stanford, ``{A bound on chaos},'' \href{http://dx.doi.org/10.1007/JHEP08(2016)106}{{\em JHEP} {\bfseries 08} (2016) 106}, \href{http://arxiv.org/abs/1503.01409}{{\ttfamily arXiv:1503.01409 [hep-th]}}.

\bibitem{Dittrich:2016hvj}
B.~Dittrich, P.~A. H\"ohn, T.~A. Koslowski, and M.~I. Nelson, ``{Can chaos be observed in quantum gravity?},'' \href{http://dx.doi.org/10.1016/j.physletb.2017.02.038}{{\em Phys. Lett. B} {\bfseries 769} (2017) 554--560}, \href{http://arxiv.org/abs/1602.03237}{{\ttfamily arXiv:1602.03237 [gr-qc]}}.

\bibitem{Turiaci:2016cvo}
G.~Turiaci and H.~Verlinde, ``{On CFT and Quantum Chaos},'' \href{http://dx.doi.org/10.1007/JHEP12(2016)110}{{\em JHEP} {\bfseries 12} (2016) 110}, \href{http://arxiv.org/abs/1603.03020}{{\ttfamily arXiv:1603.03020 [hep-th]}}.

\bibitem{Polchinski:2016xgd}
J.~Polchinski and V.~Rosenhaus, ``{The Spectrum in the Sachdev-Ye-Kitaev Model},'' \href{http://dx.doi.org/10.1007/JHEP04(2016)001}{{\em JHEP} {\bfseries 04} (2016) 001}, \href{http://arxiv.org/abs/1601.06768}{{\ttfamily arXiv:1601.06768 [hep-th]}}.

\bibitem{Jensen:2016pah}
K.~Jensen, ``{Chaos in AdS$_2$ Holography},'' \href{http://dx.doi.org/10.1103/PhysRevLett.117.111601}{{\em Phys. Rev. Lett.} {\bfseries 117} no.~11, (2016) 111601}, \href{http://arxiv.org/abs/1605.06098}{{\ttfamily arXiv:1605.06098 [hep-th]}}.

\bibitem{Stanford:2019vob}
D.~Stanford and E.~Witten, ``{JT gravity and the ensembles of random matrix theory},'' \href{http://dx.doi.org/10.4310/ATMP.2020.v24.n6.a4}{{\em Adv. Theor. Math. Phys.} {\bfseries 24} no.~6, (2020) 1475--1680}, \href{http://arxiv.org/abs/1907.03363}{{\ttfamily arXiv:1907.03363 [hep-th]}}.

\bibitem{Hashimoto:2021afd}
K.~Hashimoto, K.~Murata, N.~Tanahashi, and R.~Watanabe, ``{Bound on energy dependence of chaos},'' \href{http://dx.doi.org/10.1103/PhysRevD.106.126010}{{\em Phys. Rev. D} {\bfseries 106} no.~12, (2022) 126010}, \href{http://arxiv.org/abs/2112.11163}{{\ttfamily arXiv:2112.11163 [hep-th]}}.

\bibitem{Weber:2024ieq}
T.~Weber, J.~Tall, F.~Haneder, J.~D. Urbina, and K.~Richter, ``{Unorientable topological gravity and orthogonal random matrix universality},'' \href{http://dx.doi.org/10.1007/JHEP07(2024)267}{{\em JHEP} {\bfseries 07} (2024) 267}, \href{http://arxiv.org/abs/2405.17177}{{\ttfamily arXiv:2405.17177 [hep-th]}}.

\bibitem{Brahma:2024sie}
S.~Brahma, L.~Hackl, M.~Hassan, and X.~Luo, ``{Finite complexity of the ER=EPR state in de Sitter},'' \href{http://arxiv.org/abs/2409.13932}{{\ttfamily arXiv:2409.13932 [hep-th]}}.

\bibitem{DeFalco:2020yys}
V.~De~Falco and W.~Borrelli, ``{Detection of chaos in the general relativistic Poynting-Robertson effect: Kerr equatorial plane},'' \href{http://dx.doi.org/10.1103/PhysRevD.103.064014}{{\em Phys. Rev. D} {\bfseries 103} no.~6, (2021) 064014}, \href{http://arxiv.org/abs/2001.04979}{{\ttfamily arXiv:2001.04979 [gr-qc]}}.

\bibitem{DeFalco:2021uak}
V.~De~Falco and W.~Borrelli, ``{Timescales of the chaos onset in the general relativistic Poynting-Robertson effect},'' \href{http://arxiv.org/abs/2105.00965}{{\ttfamily arXiv:2105.00965 [gr-qc]}}.

\bibitem{Guhr:1997ve}
T.~Guhr, A.~Muller-Groeling, and H.~A. Weidenmuller, ``{Random matrix theories in quantum physics: Common concepts},'' \href{http://dx.doi.org/10.1016/S0370-1573(97)00088-4}{{\em Phys. Rept.} {\bfseries 299} (1998) 189--425}, \href{http://arxiv.org/abs/cond-mat/9707301}{{\ttfamily arXiv:cond-mat/9707301}}.

\bibitem{larkin1969quasiclassical}
A.~I. Larkin and Y.~N. Ovchinnikov, ``Quasiclassical method in the theory of superconductivity,'' {\em Sov Phys JETP} {\bfseries 28} no.~6, (1969) 1200--1205.

\bibitem{PhysRevE.86.010102}
L.~F. Santos, A.~Polkovnikov, and M.~Rigol, ``Weak and strong typicality in quantum systems,'' \href{http://dx.doi.org/10.1103/PhysRevE.86.010102}{{\em Phys. Rev. E} {\bfseries 86} (Jul, 2012) 010102}. \url{https://link.aps.org/doi/10.1103/PhysRevE.86.010102}.

\bibitem{PhysRevE.87.042135}
J.~M. Deutsch, H.~Li, and A.~Sharma, ``Microscopic origin of thermodynamic entropy in isolated systems,'' \href{http://dx.doi.org/10.1103/PhysRevE.87.042135}{{\em Phys. Rev. E} {\bfseries 87} (Apr, 2013) 042135}. \url{https://link.aps.org/doi/10.1103/PhysRevE.87.042135}.

\bibitem{Srednicki:1994mfb}
M.~Srednicki, ``{Chaos and Quantum Thermalization},'' \href{http://dx.doi.org/10.1103/PhysRevE.50.888}{{\em Phys. Rev. E} {\bfseries 50} (3, 1994) }, \href{http://arxiv.org/abs/cond-mat/9403051}{{\ttfamily arXiv:cond-mat/9403051}}.

\bibitem{berry2001chaos}
M.~V. Berry, ``Chaos and the semiclassical limit of quantum mechanics (is the moon there when somebody looks?),'' {\em Quantum Mechanics: Scientific perspectives on divine action} {\bfseries 41} (2001) 56.

\bibitem{Cornish:1996hx}
N.~J. Cornish and J.~J. Levin, ``{The Mixmaster universe: A Chaotic Farey tale},'' \href{http://dx.doi.org/10.1103/PhysRevD.55.7489}{{\em Phys. Rev. D} {\bfseries 55} (1997) 7489--7510}, \href{http://arxiv.org/abs/gr-qc/9612066}{{\ttfamily arXiv:gr-qc/9612066}}.

\bibitem{Bojowald:2023fas}
M.~Bojowald, D.~Brizuela, P.~Calizaya~Cabrera, and S.~F. Uria, ``{Chaotic behavior of the Bianchi IX model under the influence of quantum effects},'' \href{http://dx.doi.org/10.1103/PhysRevD.109.044038}{{\em Phys. Rev. D} {\bfseries 109} no.~4, (2024) 044038}, \href{http://arxiv.org/abs/2307.00063}{{\ttfamily arXiv:2307.00063 [gr-qc]}}.

\bibitem{Nosaka:2018iat}
T.~Nosaka, D.~Rosa, and J.~Yoon, ``{The Thouless time for mass-deformed SYK},'' \href{http://dx.doi.org/10.1007/JHEP09(2018)041}{{\em JHEP} {\bfseries 09} (2018) 041}, \href{http://arxiv.org/abs/1804.09934}{{\ttfamily arXiv:1804.09934 [hep-th]}}.

\bibitem{Hashimoto:2017oit}
K.~Hashimoto, K.~Murata, and R.~Yoshii, ``{Out-of-time-order correlators in quantum mechanics},'' \href{http://dx.doi.org/10.1007/JHEP10(2017)138}{{\em JHEP} {\bfseries 10} (2017) 138}, \href{http://arxiv.org/abs/1703.09435}{{\ttfamily arXiv:1703.09435 [hep-th]}}.

\bibitem{bhattacharyya2021multi}
A.~Bhattacharyya, W.~Chemissany, S.~S. Haque, J.~Murugan, and B.~Yan, ``The multi-faceted inverted harmonic oscillator: Chaos and complexity,'' {\em SciPost Physics Core} {\bfseries 4} no.~1, (2021) 002.

\bibitem{Bhattacharyya:2020art}
A.~Bhattacharyya, W.~Chemissany, S.~S. Haque, J.~Murugan, and B.~Yan, ``{The Multi-faceted Inverted Harmonic Oscillator: Chaos and Complexity},'' \href{http://dx.doi.org/10.21468/SciPostPhysCore.4.1.002}{{\em SciPost Phys. Core} {\bfseries 4} (2021) 002}, \href{http://arxiv.org/abs/2007.01232}{{\ttfamily arXiv:2007.01232 [hep-th]}}.

\bibitem{Bhattacharyya:2021cwf}
A.~Bhattacharyya, ``{Circuit complexity and (some of) its applications},'' \href{http://dx.doi.org/10.1142/S0218301321300058}{{\em Int. J. Mod. Phys. E} {\bfseries 30} no.~07, (2021) 2130005}.

\bibitem{Bhattacharyya:2019txx}
A.~Bhattacharyya, W.~Chemissany, S.~Shajidul~Haque, and B.~Yan, ``{Towards the web of quantum chaos diagnostics},'' \href{http://dx.doi.org/10.1140/epjc/s10052-022-10035-3}{{\em Eur. Phys. J. C} {\bfseries 82} no.~1, (2022) 87}, \href{http://arxiv.org/abs/1909.01894}{{\ttfamily arXiv:1909.01894 [hep-th]}}.

\bibitem{Balasubramanian:2019wgd}
V.~Balasubramanian, M.~Decross, A.~Kar, and O.~Parrikar, ``{Quantum Complexity of Time Evolution with Chaotic Hamiltonians},'' \href{http://dx.doi.org/10.1007/JHEP01(2020)134}{{\em JHEP} {\bfseries 01} (2020) 134}, \href{http://arxiv.org/abs/1905.05765}{{\ttfamily arXiv:1905.05765 [hep-th]}}.

\bibitem{Parker:2018yvk}
D.~E. Parker, X.~Cao, A.~Avdoshkin, T.~Scaffidi, and E.~Altman, ``{A Universal Operator Growth Hypothesis},'' \href{http://dx.doi.org/10.1103/PhysRevX.9.041017}{{\em Phys. Rev. X} {\bfseries 9} no.~4, (2019) 041017}, \href{http://arxiv.org/abs/1812.08657}{{\ttfamily arXiv:1812.08657 [cond-mat.stat-mech]}}.

\bibitem{Nandy:2024htc}
P.~Nandy, A.~S. Matsoukas-Roubeas, P.~Mart\'\i{}nez-Azcona, A.~Dymarsky, and A.~del Campo, ``{Quantum Dynamics in Krylov Space: Methods and Applications},'' \href{http://arxiv.org/abs/2405.09628}{{\ttfamily arXiv:2405.09628 [quant-ph]}}.

\bibitem{Brown:2016wib}
A.~R. Brown, L.~Susskind, and Y.~Zhao, ``{Quantum Complexity and Negative Curvature},'' \href{http://dx.doi.org/10.1103/PhysRevD.95.045010}{{\em Phys. Rev. D} {\bfseries 95} no.~4, (2017) 045010}, \href{http://arxiv.org/abs/1608.02612}{{\ttfamily arXiv:1608.02612 [hep-th]}}.

\bibitem{Brown:2017jil}
A.~R. Brown and L.~Susskind, ``{Second law of quantum complexity},'' \href{http://dx.doi.org/10.1103/PhysRevD.97.086015}{{\em Phys. Rev. D} {\bfseries 97} no.~8, (2018) 086015}, \href{http://arxiv.org/abs/1701.01107}{{\ttfamily arXiv:1701.01107 [hep-th]}}.

\bibitem{Auzzi:2020idm}
R.~Auzzi, S.~Baiguera, G.~B. De~Luca, A.~Legramandi, G.~Nardelli, and N.~Zenoni, ``{Geometry of quantum complexity},'' \href{http://dx.doi.org/10.1103/PhysRevD.103.106021}{{\em Phys. Rev. D} {\bfseries 103} no.~10, (2021) 106021}, \href{http://arxiv.org/abs/2011.07601}{{\ttfamily arXiv:2011.07601 [hep-th]}}.

\bibitem{https://doi.org/10.48550/arxiv.quant-ph/0502070}
M.~A. Nielsen, ``A geometric approach to quantum circuit lower bounds,'' 2005.
\newblock \url{https://arxiv.org/abs/quant-ph/0502070}.

\bibitem{Nielsen_2006}
M.~A. Nielsen, M.~R. Dowling, M.~Gu, and A.~C. Doherty, ``Quantum computation as geometry,'' \href{http://dx.doi.org/10.1126/science.1121541}{{\em Science} {\bfseries 311} no.~5764, (Feb, 2006) 1133--1135}. \url{https://doi.org/10.1126%2Fscience.1121541}.

\bibitem{https://doi.org/10.48550/arxiv.quant-ph/0701004}
M.~R. Dowling and M.~A. Nielsen, ``The geometry of quantum computation,'' 2007.
\newblock \url{https://arxiv.org/abs/quant-ph/0701004}.

\bibitem{Jefferson:2017sdb}
R.~Jefferson and R.~C. Myers, ``{Circuit complexity in quantum field theory},'' \href{http://dx.doi.org/10.1007/JHEP10(2017)107}{{\em JHEP} {\bfseries 10} (2017) 107}, \href{http://arxiv.org/abs/1707.08570}{{\ttfamily arXiv:1707.08570 [hep-th]}}.

\bibitem{Bhattacharyya:2019kvj}
A.~Bhattacharyya, P.~Nandy, and A.~Sinha, ``{Renormalized Circuit Complexity},'' \href{http://dx.doi.org/10.1103/PhysRevLett.124.101602}{{\em Phys. Rev. Lett.} {\bfseries 124} no.~10, (2020) 101602}, \href{http://arxiv.org/abs/1907.08223}{{\ttfamily arXiv:1907.08223 [hep-th]}}.

\bibitem{Ali:2019zcj}
T.~Ali, A.~Bhattacharyya, S.~S. Haque, E.~H. Kim, N.~Moynihan, and J.~Murugan, ``{Chaos and Complexity in Quantum Mechanics},'' \href{http://dx.doi.org/10.1103/PhysRevD.101.026021}{{\em Phys. Rev. D} {\bfseries 101} no.~2, (2020) 026021}, \href{http://arxiv.org/abs/1905.13534}{{\ttfamily arXiv:1905.13534 [hep-th]}}.

\bibitem{Yang:2019iav}
R.-Q. Yang and K.-Y. Kim, ``{Time evolution of the complexity in chaotic systems: a concrete example},'' \href{http://dx.doi.org/10.1007/JHEP05(2020)045}{{\em JHEP} {\bfseries 05} (2020) 045}, \href{http://arxiv.org/abs/1906.02052}{{\ttfamily arXiv:1906.02052 [hep-th]}}.

\bibitem{Balasubramanian:2021mxo}
V.~Balasubramanian, M.~DeCross, A.~Kar, Y.~C. Li, and O.~Parrikar, ``{Complexity growth in integrable and chaotic models},'' \href{http://dx.doi.org/10.1007/JHEP07(2021)011}{{\em JHEP} {\bfseries 07} (2021) 011}, \href{http://arxiv.org/abs/2101.02209}{{\ttfamily arXiv:2101.02209 [hep-th]}}.

\bibitem{Bhattacharyya:2020iic}
A.~Bhattacharyya, S.~S. Haque, and E.~H. Kim, ``{Complexity from the reduced density matrix: a new diagnostic for chaos},'' \href{http://dx.doi.org/10.1007/JHEP10(2021)028}{{\em JHEP} {\bfseries 10} (2021) 028}, \href{http://arxiv.org/abs/2011.04705}{{\ttfamily arXiv:2011.04705 [hep-th]}}.

\bibitem{Bhargava:2020fhl}
P.~Bhargava, S.~Choudhury, S.~Chowdhury, A.~Mishara, S.~P. Selvam, S.~Panda, and G.~D. Pasquino, ``{Quantum aspects of chaos and complexity from bouncing cosmology: A study with two-mode single field squeezed state formalism},'' \href{http://dx.doi.org/10.21468/SciPostPhysCore.4.4.026}{{\em SciPost Phys. Core} {\bfseries 4} (2021) 026}, \href{http://arxiv.org/abs/2009.03893}{{\ttfamily arXiv:2009.03893 [hep-th]}}.

\bibitem{Bhattacharyya:2023grv}
A.~Bhattacharyya, S.~S. Haque, G.~Jafari, J.~Murugan, and D.~Rapotu, ``{Krylov complexity and spectral form factor for noisy random matrix models},'' \href{http://dx.doi.org/10.1007/JHEP10(2023)157}{{\em JHEP} {\bfseries 10} (2023) 157}, \href{http://arxiv.org/abs/2307.15495}{{\ttfamily arXiv:2307.15495 [hep-th]}}.

\bibitem{Caputa:2021sib}
P.~Caputa, J.~M. Magan, and D.~Patramanis, ``{Geometry of Krylov complexity},'' \href{http://dx.doi.org/10.1103/PhysRevResearch.4.013041}{{\em Phys. Rev. Res.} {\bfseries 4} no.~1, (2022) 013041}, \href{http://arxiv.org/abs/2109.03824}{{\ttfamily arXiv:2109.03824 [hep-th]}}.

\bibitem{viswanath1994recursion}
V.~Viswanath and G.~M{\"u}ller, {\em The Recursion Method: Application to Many Body Dynamics}.
\newblock Lecture Notes in Physics Monographs. Springer Berlin Heidelberg, 1994.
\newblock \url{https://books.google.co.in/books?id=X2Ug4w17rnMC}.

\bibitem{Dymarsky:2019elm}
A.~Dymarsky and A.~Gorsky, ``{Quantum chaos as delocalization in Krylov space},'' \href{http://dx.doi.org/10.1103/PhysRevB.102.085137}{{\em Phys. Rev. B} {\bfseries 102} no.~8, (2020) 085137}, \href{http://arxiv.org/abs/1912.12227}{{\ttfamily arXiv:1912.12227 [cond-mat.stat-mech]}}.

\bibitem{Balasubramanian:2022dnj}
V.~Balasubramanian, J.~M. Magan, and Q.~Wu, ``{Tridiagonalizing random matrices},'' \href{http://dx.doi.org/10.1103/PhysRevD.107.126001}{{\em Phys. Rev. D} {\bfseries 107} no.~12, (2023) 126001}, \href{http://arxiv.org/abs/2208.08452}{{\ttfamily arXiv:2208.08452 [hep-th]}}.

\bibitem{Hashimoto:2023swv}
K.~Hashimoto, K.~Murata, N.~Tanahashi, and R.~Watanabe, ``{Krylov complexity and chaos in quantum mechanics},'' \href{http://dx.doi.org/10.1007/JHEP11(2023)040}{{\em JHEP} {\bfseries 11} (2023) 040}, \href{http://arxiv.org/abs/2305.16669}{{\ttfamily arXiv:2305.16669 [hep-th]}}.

\bibitem{Bhattacharyya:2023dhp}
A.~Bhattacharyya, D.~Ghosh, and P.~Nandi, ``{Operator growth and Krylov complexity in Bose-Hubbard model},'' \href{http://dx.doi.org/10.1007/JHEP12(2023)112}{{\em JHEP} {\bfseries 12} (2023) 112}, \href{http://arxiv.org/abs/2306.05542}{{\ttfamily arXiv:2306.05542 [hep-th]}}.

\bibitem{Rabinovici:2022beu}
E.~Rabinovici, A.~S\'anchez-Garrido, R.~Shir, and J.~Sonner, ``{Krylov complexity from integrability to chaos},'' \href{http://dx.doi.org/10.1007/JHEP07(2022)151}{{\em JHEP} {\bfseries 07} (2022) 151}, \href{http://arxiv.org/abs/2207.07701}{{\ttfamily arXiv:2207.07701 [hep-th]}}.

\bibitem{Chapman:2024pdw}
S.~Chapman, S.~Demulder, D.~A. Galante, S.~U. Sheorey, and O.~Shoval, ``{Krylov complexity and chaos in deformed SYK models},'' \href{http://arxiv.org/abs/2407.09604}{{\ttfamily arXiv:2407.09604 [hep-th]}}.

\bibitem{Chapman:2017rqy}
S.~Chapman, M.~P. Heller, H.~Marrochio, and F.~Pastawski, ``{Toward a Definition of Complexity for Quantum Field Theory States},'' \href{http://dx.doi.org/10.1103/PhysRevLett.120.121602}{{\em Phys. Rev. Lett.} {\bfseries 120} no.~12, (2018) 121602}, \href{http://arxiv.org/abs/1707.08582}{{\ttfamily arXiv:1707.08582 [hep-th]}}.

\bibitem{Khan:2018rzm}
R.~Khan, C.~Krishnan, and S.~Sharma, ``{Circuit Complexity in Fermionic Field Theory},'' \href{http://dx.doi.org/10.1103/PhysRevD.98.126001}{{\em Phys. Rev. D} {\bfseries 98} no.~12, (2018) 126001}, \href{http://arxiv.org/abs/1801.07620}{{\ttfamily arXiv:1801.07620 [hep-th]}}.

\bibitem{Hackl:2018ptj}
L.~Hackl and R.~C. Myers, ``{Circuit complexity for free fermions},'' \href{http://dx.doi.org/10.1007/JHEP07(2018)139}{{\em JHEP} {\bfseries 07} (2018) 139}, \href{http://arxiv.org/abs/1803.10638}{{\ttfamily arXiv:1803.10638 [hep-th]}}.

\bibitem{Guo:2018kzl}
M.~Guo, J.~Hernandez, R.~C. Myers, and S.-M. Ruan, ``{Circuit Complexity for Coherent States},'' \href{http://dx.doi.org/10.1007/JHEP10(2018)011}{{\em JHEP} {\bfseries 10} (2018) 011}, \href{http://arxiv.org/abs/1807.07677}{{\ttfamily arXiv:1807.07677 [hep-th]}}.

\bibitem{Bhattacharyya:2018bbv}
A.~Bhattacharyya, A.~Shekar, and A.~Sinha, ``{Circuit complexity in interacting QFTs and RG flows},'' \href{http://dx.doi.org/10.1007/JHEP10(2018)140}{{\em JHEP} {\bfseries 10} (2018) 140}, \href{http://arxiv.org/abs/1808.03105}{{\ttfamily arXiv:1808.03105 [hep-th]}}.

\bibitem{Ali:2018fcz}
T.~Ali, A.~Bhattacharyya, S.~Shajidul~Haque, E.~H. Kim, and N.~Moynihan, ``{Time Evolution of Complexity: A Critique of Three Methods},'' \href{http://dx.doi.org/10.1007/JHEP04(2019)087}{{\em JHEP} {\bfseries 04} (2019) 087}, \href{http://arxiv.org/abs/1810.02734}{{\ttfamily arXiv:1810.02734 [hep-th]}}.

\bibitem{Chapman:2018hou}
S.~Chapman, J.~Eisert, L.~Hackl, M.~P. Heller, R.~Jefferson, H.~Marrochio, and R.~C. Myers, ``{Complexity and entanglement for thermofield double states},'' \href{http://dx.doi.org/10.21468/SciPostPhys.6.3.034}{{\em SciPost Phys.} {\bfseries 6} no.~3, (2019) 034}, \href{http://arxiv.org/abs/1810.05151}{{\ttfamily arXiv:1810.05151 [hep-th]}}.

\bibitem{Xu:2019lhc}
T.~Xu, T.~Scaffidi, and X.~Cao, ``{Does scrambling equal chaos?},'' \href{http://dx.doi.org/10.1103/PhysRevLett.124.140602}{{\em Phys. Rev. Lett.} {\bfseries 124} no.~14, (2020) 140602}, \href{http://arxiv.org/abs/1912.11063}{{\ttfamily arXiv:1912.11063 [cond-mat.stat-mech]}}.

\bibitem{Chowdhury:2023iwg}
S.~Chowdhury, M.~Bojowald, and J.~Mielczarek, ``{Geometric quantum complexity of bosonic oscillator systems},'' \href{http://dx.doi.org/10.1007/JHEP10(2024)048}{{\em JHEP} {\bfseries 10} (2024) 048}, \href{http://arxiv.org/abs/2307.13736}{{\ttfamily arXiv:2307.13736 [quant-ph]}}.

\bibitem{Haque:2024ldr}
S.~S. Haque, G.~Jafari, and B.~Underwood, ``{Universal Early-Time Growth in Quantum Circuit Complexity},'' \href{http://arxiv.org/abs/2406.12990}{{\ttfamily arXiv:2406.12990 [hep-th]}}.

\bibitem{Caputa:2018kdj}
P.~Caputa and J.~M. Magan, ``{Quantum Computation as Gravity},'' \href{http://dx.doi.org/10.1103/PhysRevLett.122.231302}{{\em Phys. Rev. Lett.} {\bfseries 122} no.~23, (2019) 231302}, \href{http://arxiv.org/abs/1807.04422}{{\ttfamily arXiv:1807.04422 [hep-th]}}.

\bibitem{Erdmenger:2020sup}
J.~Erdmenger, M.~Gerbershagen, and A.-L. Weigel, ``{Complexity measures from geometric actions on Virasoro and Kac-Moody orbits},'' \href{http://dx.doi.org/10.1007/JHEP11(2020)003}{{\em JHEP} {\bfseries 11} (2020) 003}, \href{http://arxiv.org/abs/2004.03619}{{\ttfamily arXiv:2004.03619 [hep-th]}}.

\bibitem{Bhattacharyya:2022ren}
A.~Bhattacharyya, G.~Katoch, and S.~R. Roy, ``{Complexity of warped conformal field theory},'' \href{http://dx.doi.org/10.1140/epjc/s10052-023-11212-8}{{\em Eur. Phys. J. C} {\bfseries 83} no.~1, (2023) 33}, \href{http://arxiv.org/abs/2202.09350}{{\ttfamily arXiv:2202.09350 [hep-th]}}.

\bibitem{Chagnet:2021uvi}
N.~Chagnet, S.~Chapman, J.~de~Boer, and C.~Zukowski, ``{Complexity for Conformal Field Theories in General Dimensions},'' \href{http://dx.doi.org/10.1103/PhysRevLett.128.051601}{{\em Phys. Rev. Lett.} {\bfseries 128} no.~5, (2022) 051601}, \href{http://arxiv.org/abs/2103.06920}{{\ttfamily arXiv:2103.06920 [hep-th]}}.

\bibitem{Bhattacharyya:2023sjr}
A.~Bhattacharyya and P.~Nandi, ``{Circuit complexity for Carrollian Conformal (BMS) field theories},'' \href{http://dx.doi.org/10.1007/JHEP07(2023)105}{{\em JHEP} {\bfseries 07} (2023) 105}, \href{http://arxiv.org/abs/2301.12845}{{\ttfamily arXiv:2301.12845 [hep-th]}}.

\bibitem{Chowdhury:2024ufv}
S.~Chowdhury, M.~Bojowald, and J.~Mielczarek, ``{Upper bounds on quantum complexity of time-dependent oscillators},'' \href{http://arxiv.org/abs/2407.01677}{{\ttfamily arXiv:2407.01677 [quant-ph]}}.

\bibitem{pullen1981comparison}
R.~Pullen and A.~Edmonds, ``Comparison of classical and quantum spectra for a totally bound potential,'' {\em Journal of Physics A: Mathematical and General} {\bfseries 14} no.~12, (1981) L477.

\bibitem{PhysRevLett.69.1477}
E.~B. Bogomolny, B.~Georgeot, M.-J. Giannoni, and C.~Schmit, ``Chaotic billiards generated by arithmetic groups,'' \href{http://dx.doi.org/10.1103/PhysRevLett.69.1477}{{\em Phys. Rev. Lett.} {\bfseries 69} (Sep, 1992) 1477--1480}. \url{https://link.aps.org/doi/10.1103/PhysRevLett.69.1477}.

\bibitem{mehta2004random}
M.~Mehta, {\em Random Matrices}.
\newblock Pure and Applied Mathematics. Academic Press, 2004.
\newblock \url{https://books.google.co.in/books?id=Kp3Nx03_gMwC}.

\bibitem{ullmo2014introduction}
D.~Ullmo and S.~Tomsovic, ``Introduction to quantum chaos,'' {\em Encyclopedia of Life Support Systems (EOLSS), Oxford, UK} (2014) .

\bibitem{Cotler:2016fpe}
J.~S. Cotler, G.~Gur-Ari, M.~Hanada, J.~Polchinski, P.~Saad, S.~H. Shenker, D.~Stanford, A.~Streicher, and M.~Tezuka, ``{Black Holes and Random Matrices},'' \href{http://dx.doi.org/10.1007/JHEP05(2017)118}{{\em JHEP} {\bfseries 05} (2017) 118}, \href{http://arxiv.org/abs/1611.04650}{{\ttfamily arXiv:1611.04650 [hep-th]}}. [Erratum: JHEP 09, 002 (2018)].

\bibitem{haller1984uncovering}
E.~Haller, H.~K{\"o}ppel, and L.~Cederbaum, ``Uncovering the transition from regularity to irregularity in a quantum system,'' {\em Physical review letters} {\bfseries 52} no.~19, (1984) 1665.

\bibitem{Brown:2019whu}
A.~R. Brown and L.~Susskind, ``{Complexity geometry of a single qubit},'' \href{http://dx.doi.org/10.1103/PhysRevD.100.046020}{{\em Phys. Rev. D} {\bfseries 100} no.~4, (2019) 046020}, \href{http://arxiv.org/abs/1903.12621}{{\ttfamily arXiv:1903.12621 [hep-th]}}.

\bibitem{AIF_1966__16_1_319_0}
V.~Arnold, ``Sur la g\'eom\'etrie diff\'erentielle des groupes de {Lie} de dimension infinie et ses applications \`a l'hydrodynamique des fluides parfaits,'' \href{http://dx.doi.org/10.5802/aif.233}{{\em Annales de l'Institut Fourier} {\bfseries 16} no.~1, (1966) 319--361}. \url{https://aif.centre-mersenne.org/articles/10.5802/aif.233/}.

\bibitem{Flory:2020dja}
M.~Flory and M.~P. Heller, ``{Conformal field theory complexity from Euler-Arnold equations},'' \href{http://dx.doi.org/10.1007/JHEP12(2020)091}{{\em JHEP} {\bfseries 12} (2020) 091}, \href{http://arxiv.org/abs/2007.11555}{{\ttfamily arXiv:2007.11555 [hep-th]}}.

\bibitem{Craps:2022ese}
B.~Craps, M.~De~Clerck, O.~Evnin, P.~Hacker, and M.~Pavlov, ``{Bounds on quantum evolution complexity via lattice cryptography},'' \href{http://dx.doi.org/10.21468/SciPostPhys.13.4.090}{{\em SciPost Phys.} {\bfseries 13} no.~4, (2022) 090}, \href{http://arxiv.org/abs/2202.13924}{{\ttfamily arXiv:2202.13924 [quant-ph]}}.

\bibitem{Craps:2023rur}
B.~Craps, M.~De~Clerck, O.~Evnin, and P.~Hacker, ``{Integrability and complexity in quantum spin chains},'' \href{http://arxiv.org/abs/2305.00037}{{\ttfamily arXiv:2305.00037 [quant-ph]}}.

\bibitem{Fernandez-Melgarejo:2020fzw}
J.~J. Fernandez-Melgarejo and J.~Molina-Vilaplana, ``{Non-Gaussian Entanglement Renormalization for Quantum Fields},'' \href{http://dx.doi.org/10.1007/JHEP07(2020)149}{{\em JHEP} {\bfseries 07} (2020) 149}, \href{http://arxiv.org/abs/2003.08438}{{\ttfamily arXiv:2003.08438 [hep-th]}}.

\bibitem{Bhattacharyya:2020rpy}
A.~Bhattacharyya, S.~Das, S.~Shajidul~Haque, and B.~Underwood, ``{Cosmological Complexity},'' \href{http://dx.doi.org/10.1103/PhysRevD.101.106020}{{\em Phys. Rev. D} {\bfseries 101} no.~10, (2020) 106020}, \href{http://arxiv.org/abs/2001.08664}{{\ttfamily arXiv:2001.08664 [hep-th]}}.

\bibitem{Bhattacharyya:2020kgu}
A.~Bhattacharyya, S.~Das, S.~S. Haque, and B.~Underwood, ``{Rise of cosmological complexity: Saturation of growth and chaos},'' \href{http://dx.doi.org/10.1103/PhysRevResearch.2.033273}{{\em Phys. Rev. Res.} {\bfseries 2} no.~3, (2020) 033273}, \href{http://arxiv.org/abs/2005.10854}{{\ttfamily arXiv:2005.10854 [hep-th]}}.

\end{thebibliography}\endgroup
\bibliographystyle{utphys}





\end{document}